\documentclass{Interspeech}

\interspeechcameraready

\usepackage{multirow}

\title{Multimodal Zero-Shot Framework for Deepfake Hate Speech Detection in Low-Resource Languages}
\author[affiliation={1}]{Rishabh}{Ranjan*}
\author[affiliation={2}]{Likhith}{Ayinala*}
\author[affiliation={1}]{Mayank}{Vatsa}
\author[affiliation={1}]{Richa}{Singh}

\affiliation{}{Indian Institute of Technology Jodhpur}{India}
\affiliation{}{Columbia University}{USA}
\email{\{ranjan.4@iitj.ac.in, mvatsa, richa\}@iitj.ac.in, la3073@columbia.edu}

\keywords{Hate Speech, audio deepfakes, audio classification}

\usepackage{comment}
\usepackage{hyperref}
\begin{document}

\maketitle

\begin{abstract}
This paper introduces a novel multimodal framework for hate speech detection in deepfake audio, excelling even in zero-shot scenarios. Unlike previous approaches, our method uses contrastive learning to jointly align audio and text representations across languages. We present the first benchmark dataset with 127,290 paired text and synthesized speech samples in six languages: English and five low-resource Indian languages (Hindi, Bengali, Marathi, Tamil, Telugu). Our model learns a shared semantic embedding space, enabling robust cross-lingual and cross-modal classification. Experiments on two multilingual test sets show our approach outperforms baselines, achieving accuracies of 0.819 and 0.701, and generalizes well to unseen languages. This demonstrates the advantage of combining modalities for hate speech detection in synthetic media, especially in low-resource settings where unimodal models falter. The Dataset is available at \url{https://www.iab-rubric.org/resources}.

\end{abstract}
\def\thefootnote{*}\footnotetext{These authors contributed equally to this work}\def\thefootnote{\arabic{footnote}}

\section{Introduction}

\par Social media has transformed global communication, connecting nearly 6.3 billion users and exhibiting a remarkable compound annual growth rate-driven primarily by emerging markets in Asia, such as India, China, and Indonesia. However, this digital expansion has also amplified the reach of hate speech. Online hate speech demonstrably harms society, while current moderation techniques struggle to keep pace with the fast-evolving online landscape. This issue is further exacerbated as platforms increasingly adopt multi-modal communications across diverse languages \cite{mueller2020fanning}.

%Although evidence shows that online hate speech harms society, current moderation methods lag behind the fast-paced evolution of content, especially as platforms adopt multi-modal communications in diverse languages \cite{mueller2020fanning}.

%Despite mounting evidence of its detrimental societal impact, current moderation techniques struggle to keep pace with the rapid evolution of online content, particularly as platforms increasingly integrate multi-modal communications in diverse languages \cite{mueller2020fanning}.

%\par Social media has transformed global communication, connecting nearly 6.3 billion users with a projected 26.84\% compound annual growth rate (CAGR), driven by Asian countries like India, China, and Indonesia. Despite the significant harm caused by online hate speech, moderation efforts\cite{mueller2020fanning} have not kept pace with the evolving landscape, particularly as multi-modal content in various languages becomes more prevalent.

Existing research on hate speech detection predominantly focuses on textual analysis, often overlooking the rich information contained in other modalities. Prior studies and datasets \cite{Mathew2021HateXplainAB, ousidhoum-etal-2019-multilingual, i2019multilingual, DBLP:journals/corr/abs-2006-08328} have primarily centered on conversational text, leaving audio-based modalities relatively underexplored due to technical challenges and limited datasets. In the audio domain, approaches are typically categorized into two types: cascaded systems, which extend traditional text toxicity detection by incorporating speech recognition pipelines \cite{communication2023seamlessm4tmassively}, and end-to-end systems, which classify toxicity directly from audio signals \cite{Ghosh2021DeToxyAL}. While the latter approach shows promise, it has primarily been validated on English datasets, demonstrating significant advantages over text-based models in handling out-of-domain content, as shown in studies using proprietary datasets \cite{DBLP:conf/eusipco/YousefiE21}. Recent advancements, such as MuTox \cite{DBLP:conf/acl/Costa-jussaMADH24}, have introduced scalable multilingual audio-based toxicity detectors capable of zero-shot detection across multiple languages, representing a significant leap forward in this field.

%However, most prior work in literature\cite{Mathew2021HateXplainAB, ousidhoum-etal-2019-multilingual, i2019multilingual, DBLP:journals/corr/abs-2006-08328} and available datasets focus primarily on the modality of conversational text, with other modalities of conversation ignored at large. When exploring audio-based hate-speech detection, there are either cascaded systems which extend text toxicity detection with speech recognition\cite{communication2023seamlessm4tmassively}; or end-to-end audio-based toxicity classification\cite{Ghosh2021DeToxyAL} which provides an English dataset together with end-to-end toxicity detection results. This work shows that gains of English text-less audio-based classifiers over text-based classifiers are specially relevant when applied to out-of-domain, coherently with the previous study on a non-disclosed dataset\cite{DBLP:conf/eusipco/YousefiE21}. MuTox\cite{DBLP:conf/acl/Costa-jussaMADH24} proposes a highly massive multilingual audio-based toxicity detector. Authors show that MuTox can detect zero-shot toxicity across various languages.
\begin{figure}
    \centering
    \includegraphics[width=\columnwidth]{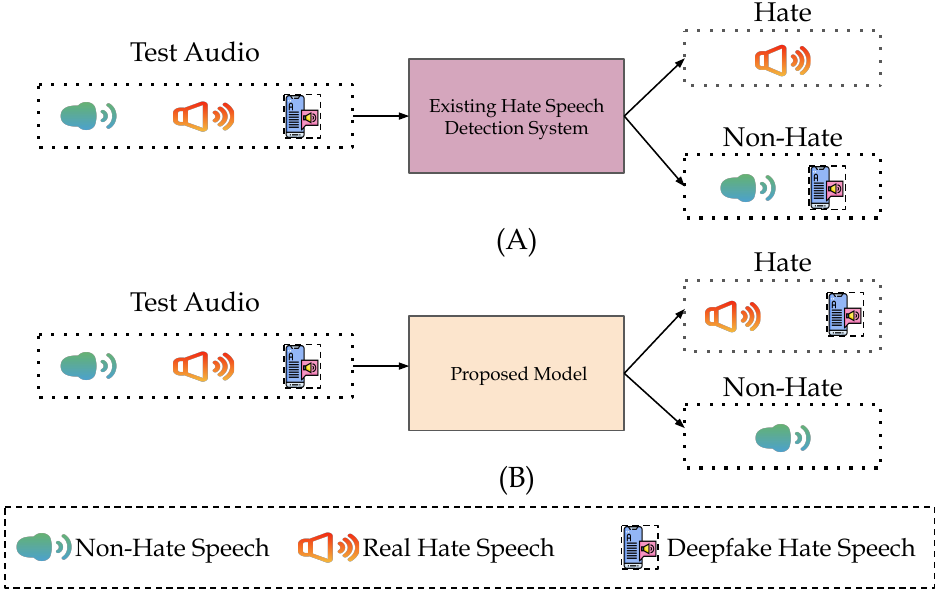}
    \caption{(A) Current hate speech detectors often misclassify deepfake-generated hate speech audio as non-hateful, exposing moderation systems to manipulation. (B) The proposed multimodal framework effectively distinguishes deepfake hate speech from genuine non-hateful content, enhancing detection accuracy and robustness.}
    \label{fig:visual_proposed}
    \vspace{-10pt}
\end{figure}

\par Despite these advancements, a critical research gap remains: the detection of hate speech in deepfake audios. The rise of synthetic hate content online has raised concerns about its potential to manipulate public discourse; however, current literature and datasets have not addressed this challenge. Traditional hate speech detection datasets primarily focus on textual content \cite{Mathew2021HateXplainAB, Bhardwaj2020HostilityDD, marreddy2022resource}, and while some efforts have expanded into audio \cite{DBLP:conf/acl/Costa-jussaMADH24, Ghosh2021DeToxyAL, adima_paper}, none have tackled the unique challenges posed by deepfake audio manipulations, such as their synthetic nature and potential for deception. Figure \ref{fig:visual_proposed} shows the limitations of the current deepfake detection system.

\par To bridge this gap, we introduce a novel dataset comprising paired audio and text samples from English and five low-resource languages-Hindi, Bengali, Marathi, and Tamil, and curated explicitly for hate speech detection in deepfake audios. These languages are considered low-resource due to limited datasets and computational tools. Alongside this dataset, we propose a multi-modal architecture that leverages state-of-the-art text and audio encoders to project heterogeneous data into a unified embedding space using contrastive loss that enhances similarity learning. This joint embedding approach not only enables robust cross-lingual and cross-modal detection but also equips our system with zero-shot capabilities, allowing it to generalize to unseen languages and modalities.

%\par Our contributions are threefold. First, we provide a multi-modal dataset addressing the gap in deepfake audio hate speech detection. Second, we introduce a novel, contrastive learning-based architecture that effectively fuses textual and acoustic features. Finally, we establish strong multi-modal baselines and demonstrate the superiority of our approach over traditional text-based detectors, particularly in zero-shot scenarios. This work lays a foundation for future research in multimodal hate speech detection, paving the way for safer digital environments through effective content moderation strategies.

Our contributions are threefold. First, we provide a multi-modal dataset addressing the gap in deepfake audio hate speech detection. Second, we introduce a novel, contrastive learning-based architecture that effectively fuses textual and acoustic features. Finally, we establish strong multi-modal baselines and demonstrate the superiority of our approach over traditional text-based detectors, particularly in zero-shot scenarios, where ``zero-shot'' refers to detecting hate speech in new languages. This work lays a foundation for future research in multimodal hate speech detection, paving the way for safer digital environments through effective content moderation strategies.

\section{Proposed Hate Audio Detector Model}

\begin{figure}[t!]
    \centering
    \includegraphics[width = 0.8\columnwidth]{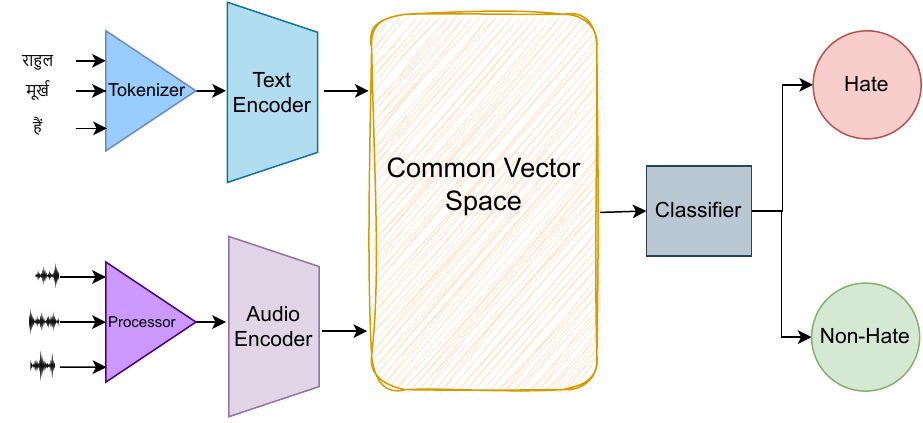}
    \caption{Proposed multimodal hate speech detection pipeline: Audio and text inputs are encoded separately and mapped into a common semantic embedding space, enabling effective cross-modal and cross-lingual classification through contrastive learning.}
    \label{fig:architecture}
    \vspace{-10pt}
\end{figure}

\par The proposed framework introduces a novel two-stage contrastive learning approach for multimodal hate speech detection, distinctly combining audio and text modalities. Unlike existing methods, we employ a unique alignment strategy in the pre-training stage, leveraging state-of-the-art encoders, SONAR for text and SeamlessM4T for audio, to project embeddings into a unified semantic space. This multimodal alignment significantly enhances zero-shot generalization capabilities, particularly in low-resource languages. Formally, our objective is to learn a classifier  $f_{\theta}$.
%The proposed framework integrates audio and text encoders and is trained using a novel two-stage contrastive learning approach. In the initial stage, the framework aligns audio and text embeddings in a manner similar to the method employed in Audclip. This is followed by a second stage that utilizes a novel supervised contrastive loss function. This loss function minimizes the distance between embeddings of audio-text pairs from the same class (Hate or Non-Hate) across different languages, while maximizing the distance between embeddings of pairs from different classes. Figure \ref{fig:architecture} shows the proposed pipeline for deepfake hate speech detection. The objective of the framework is to learn a function \( f_\theta \) that maps an audio-text pair \( (a_i, t_i) \) to a score \( s^u \). This score \( s^u \) indicates whether the pair belongs to the Hate (denoted by 0) or Non-Hate (denoted by 1) class. The task is formalized in Equation \ref{eq:task}:
\begin{equation}
    f_{\theta}: (a_i, t_i) \mapsto s_u,\quad \text{where} \quad a_i \in \mathbb{R}^{d_a}, \; t_i \in \mathbb{R}^{d_t}, \; s_u \in [0,1]
\label{eq:task}
\end{equation}
By minimizing intra-class distances and maximizing inter-class distances in the embedding space, the proposed approach effectively addresses cross-lingual and cross-modal classification challenges, significantly advancing the state-of-the-art in hate speech detection for synthetic audio.
%In the proposed hate speech audio detection framework, we employ two stages: a contrastive pre-training phase followed by a downstream phase using supervised contrastive training. %Details are elaborated below.

%The detailed methodology of the proposed framework is described as follows. First, we elaborate on the contrastive pre-training phase. Subsequently, we discuss the supervised contrastive training phase and provide an overview of the complete pipeline.

%\par The proposed framework integrates audio and text encoders and is trained using a novel two-stage contrastive learning approach. In the initial stage, the framework aligns audio and text embeddings, akin to the method employed in Audclip. This is followed by a second stage involving a novel supervised contrastive loss function. This loss function is designed to minimize the distance between embeddings of audio-text pairs from the same class (Hate or Non-Hate) across different languages while maximizing the distance between embeddings of different classes. The objective of the framework is to learn a function \( f_\theta \) that maps an audio-text pair \( (a_i, t_i) \) to a score \( s^u \). This score \( s^u \) indicates whether the pair belongs to the Hate (denoted by 0) or Non-Hate (denoted by 1) class. The task definition is formalized in Equation \ref{eq:task}:

%The detailed methodology of the proposed framework is described as follows: First, we elaborate on the contrastive pre-training phase. Subsequently, we discuss the supervised contrastive training phase and provide an overview of the complete pipeline.

\subsection{Pre-training Phase}
\noindent In developing the hate speech classification network, we implement a sophisticated pre-training approach that leverages transfer learning and contrastive learning techniques. This method enables us to exploit rich representations from existing models while fine-tuning for our target multimodal hate speech detection task. Let \(a_i\) represent the audio input and \(T_i\) represent the text input. Our network architecture comprises two primary components: a text encoder \(f_T\) and an audio encoder \(f_A\). The weights of the text encoder \(f_T\) are initialized with weights \(\theta_T\) from the SONAR encoder (i.e., \(\theta_T = \theta_{SONAR}\)), and the audio encoder \(f_A\) is initialized with weights \(\theta_A\) from SeamlessM4T (i.e., \(\theta_A = \theta_{SeamlessM4T}\)). This choice is motivated by the zero-shot capabilities of SONAR: once trained on a set of languages, the classifier head can seamlessly integrate with any compatible SONAR encoder for different languages.

\par The core of our pre-training process lies in the contrastive learning framework. We aim to create a common vector space \(\mathcal{V} \subset \mathbb{R}^{m \times d}\) where textual and audio representations can be meaningfully compared, independent of language. Here, \(m \times d\) denotes the dimensions of the output embedding space. For a batch of \(N\) samples, let \(\{(a_i, T_i, y_i)\}_{i=1}^N\) represent the audio inputs, text inputs, and their corresponding labels, respectively. The encoders produce embeddings as 
\[
e_{A_i} = f_A(a_i), \quad e_{T_i} = f_T(T_i).
\]

\par These embeddings are normalized to ensure they reside on a unit hypersphere, as shown in Equation \ref{eqn:unihyper}:
\begin{equation}
\hat{e}_{A_i} = \frac{e_{A_i}}{\lVert e_{A_i} \rVert_2}, \quad 
\hat{e}_{T_i} = \frac{e_{T_i}}{\lVert e_{T_i} \rVert_2}.
\label{eqn:unihyper}
\end{equation}

\par We compute a similarity matrix \(S \in \mathbb{R}^{N \times N}\) by taking the dot product between all pairs of audio and text embeddings, defined as \(S_{ij} = \hat{e}_{A_i}^T \hat{e}_{T_j}\). Positive and negative masks are then created based on the labels: the positive mask is defined as \(M_{pos_{ij}} = \mathbb{1}[y_i = y_j]\) and the negative mask as \(M_{neg_{ij}} = \mathbb{1}[y_i \neq y_j]\).

\par During pre-training, we optimize the contrastive loss \(\mathcal{L}\) via \(\min_{\theta_T, \theta_A} \mathcal{L}\). This process encourages the model to maximize the similarity of positive pairs while minimizing that of negative pairs in the shared embedding space \(\mathcal{V}\). The use of a common vector space facilitates cross-modal learning, leveraging the relationship 
\[
\text{sim}(f_A(a_i), f_T(T_j)) \approx \mathbb{1}[y_i = y_j],
\]
where, \(\text{sim}\) is a similarity function (e.g., cosine similarity).

\begin{figure}[t!]
    \centering
    \includegraphics[width=0.8\columnwidth]{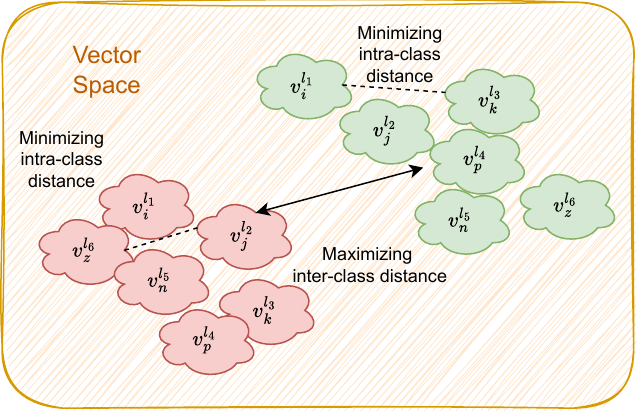}
    \caption{Illustration of contrastive training to maximize inter-class distance while reducing intra-class distance between embeddings of different languages.}
    \label{fig:downstream}
    \vspace{-10pt}
\end{figure}

\subsection{Downstream Phase}
\noindent Following the pre-training phase, we proceed to the downstream phase, where we fine-tune our model for the specific task of hate speech classification. In this phase, we train a classifier on top of our pre-trained encoders while continuing to refine their weights. We employ a novel approach that combines triplet loss and binary cross-entropy loss to enhance the model's discriminative power across different languages. Figure \ref{fig:downstream} illustrates the triplet-based contrastive learning phase.

\noindent Our model architecture in this phase consists of the pre-trained audio encoder \(f_A: \mathbb{R}^d \rightarrow \mathbb{R}^m\) and text encoder \(f_T: \mathbb{R}^{n \times d} \rightarrow \mathbb{R}^m\), along with a newly introduced classifier \(g: \mathbb{R}^{2m} \rightarrow [0, 1]\) that takes the concatenated embeddings from both encoders and outputs a probability indicating hate speech. For an input pair \((a_i, T_i)\), the classification process is defined as 
\[
e_i = [f_A(a_i); f_T(T_i)] \quad \text{and} \quad \hat{y}_i = g(e_i),
\]
where, \([;]\) denotes concatenation and \(\hat{y}_i\) is the predicted probability of hate speech. The loss functions used in this phase are crucial to our approach. We utilize a combination of triplet loss and binary cross-entropy loss. The triplet loss is designed to maximize inter-class distance while minimizing intra-class distance in the embedding space, thereby creating a language-independent vector space for hate and non-hate representations. For each language \(l\), we form triplets \((e_a, e_p, e_n)\) where \(e_a\) (anchor) is the embedding of a hate speech sample in language \(l\), \(e_p\) (positive) is the embedding of another hate speech sample in language \(l\), and \(e_n\) (negative) is the embedding of a non-hate speech sample in a different language \(l' \neq l\).

\par The total loss for the downstream phase is a weighted sum of the triplet loss, \(\mathcal{L}_{triplet}\), and the binary cross-entropy loss, \(\mathcal{L}_{BCE}\). The complete loss formulation is given by Equation \ref{eqn:triplet}:
\begin{equation}
\mathcal{L}_{total} = \alpha \mathcal{L}_{triplet} + (1-\alpha) \mathcal{L}_{BCE},
\label{eqn:triplet}
\end{equation}
where, \(\alpha \in [0, 1]\) is a hyperparameter controlling the balance between the two losses. During the downstream phase, we optimize
\[
\min_{\theta_A, \theta_T, \theta_g} \mathcal{L}_{total},
\]
where, \(\theta_A\), \(\theta_T\), and \(\theta_g\) are the parameters of the audio encoder, text encoder, and classifier, respectively. This downstream training phase refines our model's hate speech detection capabilities while maintaining robustness across different languages. The triplet loss drives the model to learn language-invariant features of hate speech, while the binary cross-entropy loss ensures accurate classification. The result is a model that can effectively identify hate speech in a multilingual context by leveraging both audio and textual information.

\begin{table}[t!]
\centering
\caption{Distribution of hate and non-hate samples in the proposed dataset.}
\label{tab:sample_table}
\resizebox{\columnwidth}{!}{%
\begin{tabular}{|lccccc|}
\hline
\multicolumn{1}{|l|}{\multirow{2}{*}{\textbf{Language}}} & \multicolumn{2}{c|}{\textbf{Hate}}                                       & \multicolumn{2}{c|}{\textbf{Non-Hate}}                                   & \multirow{2}{*}{\textbf{Total}} \\ \cline{2-5}
\multicolumn{1}{|l|}{}                                   & \multicolumn{1}{c|}{\textbf{Train}} & \multicolumn{1}{c|}{\textbf{Test}} & \multicolumn{1}{c|}{\textbf{Train}} & \multicolumn{1}{c|}{\textbf{Test}} &                                 \\ \hline
\multicolumn{1}{|l|}{English}                            & \multicolumn{1}{c|}{6251}           & \multicolumn{1}{c|}{9132}          & \multicolumn{1}{c|}{1142}           & \multicolumn{1}{c|}{782}           & 17307                           \\ \hline
\multicolumn{1}{|l|}{Hindi}                              & \multicolumn{1}{c|}{2149}           & \multicolumn{1}{c|}{2433}          & \multicolumn{1}{c|}{529}            & \multicolumn{1}{c|}{617}           & 5728                            \\ \hline
\multicolumn{1}{|l|}{Marathi}                            & \multicolumn{1}{c|}{15000}          & \multicolumn{1}{c|}{15000}         & \multicolumn{1}{c|}{1875}           & \multicolumn{1}{c|}{1875}          & 33750                           \\ \hline
\multicolumn{1}{|l|}{Tamil}                              & \multicolumn{1}{c|}{1926}           & \multicolumn{1}{c|}{1831}          & \multicolumn{1}{c|}{815}            & \multicolumn{1}{c|}{791}           & 5363                            \\ \hline
\multicolumn{1}{|l|}{Telugu}                             & \multicolumn{1}{c|}{259}            & \multicolumn{1}{c|}{24340}         & \multicolumn{1}{c|}{111}            & \multicolumn{1}{c|}{10432}         & 35142                           \\ \hline
\multicolumn{1}{|l|}{Bengali}                            & \multicolumn{1}{c|}{7000}           & \multicolumn{1}{c|}{14000}         & \multicolumn{1}{c|}{3000}           & \multicolumn{1}{c|}{600}           & 30000                           \\ \hline
\multicolumn{6}{|c|}{Total : 127290}                                                                                                                                                                                                             \\ \hline
\end{tabular}%
}
\end{table}

\section{Proposed Dataset and Experimental Details}
We introduce a novel multimodal, multilingual dataset designed to address the scarcity of resources for hate speech detection, with a focus on low-resource languages. The dataset encompasses both textual and audio modalities across six languages: English, Hindi, Telugu, Tamil, Marathi, and Bengali, chosen for their linguistic diversity and representation in hate speech.

\noindent\textbf{Data Collection and Processing:} The dataset was constructed by converting existing text corpora into audio format using state-of-the-art text-to-speech (TTS) synthesis technology, specifically Meta's Massive Multi-Lingual model \cite{abelson-et-al:scheme}, chosen for its robust multilingual capabilities. Audio samples were standardized to a 16 kHz sampling rate, consistent with common speech processing standards, and limited to 10 seconds in duration to balance information retention and efficiency.

\noindent\textbf{Dataset Composition:} The textual sources were selected from peer-reviewed collections to ensure data quality and reliability. The English subset, derived from the HateXplain dataset \cite{Mathew2021HateXplainAB}, contains 17,307 social media samples annotated via Amazon Mechanical Turk, ensuring high-quality labels for hate speech, offensive language, and neutral content. The Hindi subset comprises 5,728 samples from the Hostility Dataset \cite{Bhardwaj2020HostilityDD}. The Marathi subset, with 33,750 samples from \cite{patil2022l3cube}, is particularly significant due to its large size, addressing the scarcity of hate speech data in this language. The Telugu subset, focused on news website comments, contributes 35,142 samples from \cite{marreddy2022resource}. The Tamil subset includes 5,363 samples from the Tamil Hate Speech Project, a collaborative academic initiative annotated by linguistic experts. Lastly, the Bengali subset offers 30,000 samples from \cite{Romim2020HateSD}, sourced from YouTube comments on controversial topics, which were carefully filtered to remove irrelevant content. Table \ref{tab:sample_table} provides the detailed distribution of hate and non-hate samples across languages and train-test splits, highlighting notable variations in balance and size stemming from the differing scales of the underlying text corpora. 

\textit{To the best of our knowledge, this is the first dataset to leverage synthetically generated speech data for multilingual, multimodal hate speech detection, addressing the lack of paired text-audio resources in low-resource languages.} By combining text and audio modalities across multiple low-resource languages, this dataset enables research in cross-lingual and cross-modal hate speech detection, as well as the development of robust multimodal models for real-world applications. The dataset will be publicly available to support further research and development.

\subsection{Experimental Setup}
We focus on two primary research objectives: \textit{RO1 - Multilingual Deepfake Hate Speech Detection:} To evaluate the accuracy of the proposed dataset and model in detecting hate speech in deepfake audio across multiple languages.
\noindent \textit{RO2 - Zero-Shot Hate Speech Detection:}  To assess the zero-shot learning capabilities of the dataset and model for hate speech detection in languages unseen during training.

\noindent\textbf{Dataset Protocols:} To thoroughly evaluate zero-shot learning capabilities, we employ a cross-lingual train-test split. First, we partition each language’s data into training and testing subsets using a 70:30 ratio. Next, we define two distinct language sets: (i) \textbf{Set-A}: Marathi, Bengali, and Tamil, and (ii) \textbf{Set-B}: English, Hindi, and Telugu. These languages were selected to represent diverse linguistic families and ensure a rigorous evaluation of multilingual generalization. We conduct two experiments: (1) training on Set-A and testing on Set-B, and (2) training on Set-B and testing on Set-A. This approach allows us to assess generalization across entirely unseen languages, providing a stringent test of zero-shot learning and multilingual robustness.

\subsection{Baselines and Implementation Details} To benchmark the performance of the proposed model, we compare it against six baseline classifiers, selected for their state-of-the-art performance in hate speech detection, multilingual tasks, and multimodal learning:

\noindent \textbf{Text-Based Models:} (i) \textit{SentenceBERT} \cite{Burnap2014HateSM}: a fine-tuned BERT model with attention-based pooling, (ii) \textit{HASOC22} \cite{10.1145/3574318.3574326}: a linear-layer-based architecture optimized on multilingual hate speech data, and (iii) \textit{CNNGRU} \cite{inproceedings}: a hybrid model combining convolutional and recurrent (GRU) layers.

\noindent \textbf{Multimodal Models:} (i) \textit{ASTBERT} \cite{DBLP:journals/corr/abs-2104-01778}: which combines an Audio Spectrogram Transformer with BERT-based text embeddings, (ii) \textit{HUBERTWavBERT} \cite{DBLP:journals/corr/abs-2106-07447}: which integrates a HuBERT-based audio encoder with a BERT text encoder, and (iii) \textit{WAVELMBERT} \cite{DBLP:journals/corr/abs-2110-13900}: which merges WaveLM-based audio embeddings with a BERT text encoder.

%\noindent\textbf{Multimodal Models:} (i) \textit{ASTBERT} \cite{DBLP:journals/corr/abs-2104-01778}: Combines an Audio Spectrogram Transformer with BERT-based text embeddings, (ii) \textit{HUBERTWavBERT} \cite{DBLP:journals/corr/abs-2106-07447}: Integrates a HuBERT-based audio encoder with a BERT text encoder, and (iii) \textit{WAVELMBERT} \cite{DBLP:journals/corr/abs-2110-13900}: Merges WaveLM-based audio embeddings with a BERT text encoder.

\par All models use the multilingual BERT-uncased tokenizer and are trained for five epochs using cross-entropy loss and the Adam optimizer (learning rate = 0.0001, batch size = 32, dropout rate = 0.2). Performance is evaluated using accuracy, F1-score, and AUC-ROC metrics to ensure a comprehensive assessment of the proposed model’s capabilities.

\begin{table}[t!]
\caption{Results of the Proposed and baseline models when training and testing are on the same set.}
\label{tab:sameset_results_table}
\resizebox{\columnwidth}{!}{%
\begin{tabular}{|l|c|cc|cc|}
\hline
\multirow{2}{*}{Model} & \multirow{2}{*}{Input} & \multicolumn{2}{c|}{Set-A}          & \multicolumn{2}{c|}{Set-B}          \\ \cline{3-6} 
                       &                        & \multicolumn{1}{c|}{ACC}   & EER   & \multicolumn{1}{c|}{ACC}   & EER   \\ \hline
ASTBERT                & Text + Audio           & \multicolumn{1}{c|}{0.777} & 0.223 & \multicolumn{1}{c|}{0.669} & 0.331 \\ \hline
HUBERT                 & Text + Audio           & \multicolumn{1}{c|}{0.762} & 0.237 & \multicolumn{1}{c|}{0.635} & 0.369 \\ \hline
WAVELMBERT             & Text + Audio           & \multicolumn{1}{c|}{0.790} & 0.210 & \multicolumn{1}{c|}{0.669} & 0.331 \\ \hline
SENTENCEBERT           & Text                   & \multicolumn{1}{c|}{0.773} & 0.228 & \multicolumn{1}{c|}{0.653} & 0.346 \\ \hline
HASOC'22               & Text                   & \multicolumn{1}{c|}{0.755} & 0.244 & \multicolumn{1}{c|}{0.637} & 0.359 \\ \hline
CNN-GRU                & Text                   & \multicolumn{1}{c|}{0.738} & 0.262 & \multicolumn{1}{c|}{0.620} & 0.380 \\ \hline
Proposed               & Text + Audio           & \multicolumn{1}{c|}{0.819} & 0.181 & \multicolumn{1}{c|}{0.701} & 0.301 \\ \hline
\end{tabular}%
}
\end{table}
%\vspace{-5pt}
\section{Results and Analysis}

\noindent \textbf{RO1: Multilingual Deepfake Hate Speech Detection.} We evaluate the performance of our proposed model on two language sets: Set-A (Tamil, Marathi, Bengali) and Set-B (English, Hindi, Telugu). These experiments address RO1 by demonstrating the effectiveness of our approach for multilingual deepfake hate speech detection across diverse language groups. For Set-A (see Table \ref{tab:sameset_results_table}), our proposed model achieves an accuracy of 0.819 and an Equal Error Rate (EER) of 0.181, highlighting its strong performance. Among the baseline methods, WAVELMBERT performs best with an accuracy of 0.790, followed closely by ASTBERT and SENTENCEBERT. Notably, text-only models (SENTENCEBERT, HASOC'22, CNN-GRU) generally underperform compared to multimodal models, suggesting that incorporating audio features significantly enhances hate speech detection.

The results for Set-B (also shown in Table \ref{tab:sameset_results_table}) indicate that our model maintains robust multilingual capabilities, achieving an accuracy of 0.701 and an EER of 0.301. ASTBERT and WAVELMBERT achieve comparable performance, with an accuracy of 0.669. Interestingly, the performance gap between multimodal and text-only models is less pronounced in Set-B than in Set-A. This may suggest that audio features are less informative for these languages or that the models face challenges in generalizing audio features across linguistically distant languages. Overall, our proposed model demonstrates the most consistent performance across all languages and both sets, indicating superior cross-lingual generalization capabilities. Additionally, WAVELMBERT and ASTBERT exhibit strong and consistent performance, making them viable alternatives for multilingual hate speech detection tasks.

\begin{table}[t!]
\caption{Results for the cross-subset evaluation. The results highlight the Zero-Shot capabilities of our Proposed model.}
\label{tab:cross_results}
\resizebox{\columnwidth}{!}{%
\begin{tabular}{|l|l|ll|ll|}
\hline
\multirow{2}{*}{Model} & \multirow{2}{*}{Input} & \multicolumn{2}{l|}{Train Set-A, Eval Set-B} & \multicolumn{2}{l|}{Train Set-B, Eval Set-A} \\ \cline{3-6} 
                       &                        & \multicolumn{1}{l|}{ACC}        & EER        & \multicolumn{1}{l|}{ACC}        & EER        \\ \hline
ASTBERT                & Text + Audio           & \multicolumn{1}{l|}{0.602}      & 0.398      & \multicolumn{1}{l|}{0.750}      & 0.250      \\ \hline
HUBERT                 & Text + Audio           & \multicolumn{1}{l|}{0.572}      & 0.428      & \multicolumn{1}{l|}{0.723}      & 0.277      \\ \hline
WAVELMBERT             & Text + Audio           & \multicolumn{1}{l|}{0.596}      & 0.403      & \multicolumn{1}{l|}{0.752}      & 0.248      \\ \hline
SENTENCEBERT           & Text                   & \multicolumn{1}{l|}{0.595}      & 0.406      & \multicolumn{1}{l|}{0.746}      & 0.254      \\ \hline
HASOC'22               & Text                   & \multicolumn{1}{l|}{0.581}      & 0.421      & \multicolumn{1}{l|}{0.731}      & 0.270      \\ \hline
CNN-GRU                & Text                   & \multicolumn{1}{l|}{0.560}      & 0.442      & \multicolumn{1}{l|}{0.707}      & 0.293      \\ \hline
PROPOSED               & Text + Audio           & \multicolumn{1}{l|}{0.625}      & 0.373      & \multicolumn{1}{l|}{0.786}      & 0.214      \\ \hline
\end{tabular}%
}
\end{table}
%The proposed model demonstrates the most consistent performance across all languages and both sets, indicating superior cross-lingual generalization capabilities. WAVELMBERT and ASTBERT also show strong and consistent performance, making them robust alternatives for multilingual hate speech detection. Some languages, like Marathi, show high performance across multiple models, while others, like Telugu, consistently yield lower accuracies. This suggests that language-specific characteristics play a crucial role in model performance and that future work should address these disparities.\\

\noindent \textbf{RO2: Zero-Shot Hate Speech Detection.} We conducted cross-subset experiments to evaluate the models' ability to generalize across different language groups, thereby assessing the zero-shot capabilities of the proposed model. Table \ref{tab:cross_results} presents the results for two scenarios: (1) training on Set-A (Tamil, Marathi, Bengali) and evaluating on Set-B (English, Hindi, Telugu), and (2) training on Set-B and evaluating on Set-A. When training on Set-A and evaluating on Set-B, our proposed model achieves the highest accuracy of 0.625 and the lowest Equal Error Rate (EER) of 0.373, outperforming all baselines. ASTBERT follows with an accuracy of 0.602, while other models show performance ranging from 0.560 to 0.596. These results highlight the challenge of generalizing from the languages in Set-A to those in Set-B, potentially due to linguistic differences and data characteristics.

Interestingly, when training on Set-B and evaluating on Set-A, all models demonstrate a significant improvement. Our proposed model achieves the highest accuracy of 0.786 and an EER of 0.214, with WAVELMBERT and ASTBERT following closely at 0.752 and 0.750, respectively. This substantial improvement suggests that the models can better generalize from the languages in Set-B to those in Set-A. The presence of English in Set-B - a widely used language with abundant training data - may have contributed to more robust feature learning, facilitating cross-lingual transfer. The consistent outperformance of multimodal models in these scenarios further indicates that incorporating audio features provides valuable information that generalizes well across language sets. This emphasizes the importance of multimodal approaches in multilingual hate speech detection tasks. Overall, the superior performance of our proposed model in both cross-subset experiments demonstrates its robust zero-shot capabilities and highlights the need for models that can effectively generalize across diverse language groups, particularly when transferring from less commonly spoken to more widely spoken languages.

\section{Conclusion}
This paper presents a novel zero-shot hate speech detection approach for audio, addressing challenges posed by multimodal content in low-resource languages. Our contributions include: (1) a framework that integrates audio and text modalities using contrastive learning; (2) the first comprehensive multimodal dataset for deepfake hate speech detection, covering English and five low-resource Indian languages; and (3) a two-stage training process combining contrastive pre-training and supervised contrastive learning. Our proposed model consistently outperforms existing baselines, with multimodal approaches demonstrating superiority over text-only models in both same-language and cross-language scenarios. This work highlights the value of incorporating audio features alongside text for robust hate speech detection in low-resource contexts. Future work will expand the dataset with diverse deepfake audio samples featuring varied voice characteristics, accents, and emotional tones to further enhance detection capabilities.

\section{Acknowledgement}
\vspace{-2pt}
% This work is supported by ACM IKDD. Thakral received partial support from the PMRF Fellowship and Vatsa is partially supported by the Swarnajayanti Fellowship.

This research is supported by a grant from the NSM, MeitY. The authors also gratefully acknowledge the support of IndiaAI and Meta through Srijan: Centre of Excellence for Generative AI.

\bibliographystyle{IEEEtran}
\bibliography{mybib}

\end{document}